\documentstyle[epsfig]{aipproc}
\input pah0.eps
\begin{document}
 
\title{Models for Type Ia Supernovae and Evolutionary Effects with Redshift}
\author{P. H\"oflich}
\address{ Department of Astronomy, University of Texas, Austin, TX 78681, USA}

\maketitle

\begin{abstract} 
Based on detailed models for the explosions, light curves and NLTE-spectra,
evolutionary effects of Type Ia Supernovae (SNe~Ia) with redshift have been studied to evaluate
their size on cosmological time scales, how the effects can be recognized and how one may
be able to correct for them.
 
 In the first part, we summarize the current status of scenarios for Type Ia Supernovae, including the
explosion
of a Chandrasekhar mass           white dwarf ($M_{Ch}$-WD), the merging scenario and the helium-triggered explosions of
low-mass WDs. We show that
delayed detonation models  can account for the majority of observations of spectra and light curves.
  IR observations are a new and powerful tools  to constrain explosion models and
details of the flame propagation in the WD. A strong Mg II line at 1.05 $\mu m$ shows that
nuclear burning takes place at the outer, low density layers. This requires a transition from the
deflagration to the detonation regime of the nuclear burning front, or a very fast deflagration
close to the speed of sound.
We put the models into context with the empirical brightness decline relation which is
 widely applied to use SNe~Ia as yardsticks on 
cosmological distance scales. This relation can be well understood
in the framework of $M_{Ch}$-WDs as a consequence of the opacity effects in combination with the
amount of $^{56}Ni$ which determines the brightness.
 However, the amount of $^{56}Ni$
actually produced depends on a combination of free parameters such as central density and the 
chemical composition of the WD, and  the propagation of the burning front. We get a spread
of $\approx 0.4 ^m$ around the mean relation which is  larger than currently favored by
observations ($\approx 0.12^m$, \cite{riess98}) which 
 may hint of an underlying coupling of the progenitor, the accretion rates and the
propagation of the burning front. 
     
 In a second  part, we investigate the possible evolutionary effects in SNe~Ia  both
 with respect to changes in the sample of SNe~Ia and individual variations, and how they can 
 be identified. We find that evolution may produce an offset in the brightness decline relation  is
 restricted to a few tenth of a magnitude.
  The effects reveal themself by  changes in the U and UV fluxes, and in a change
   in the maximum brightness/decline relation
   by $\Delta M \approx 0.1 \times \Delta t$ where $\Delta t$ is the difference between local and
 distant SN-samples.
 According to new data \cite{aldering2000}, $\Delta t \leq 1  d$ and, thus,
  evolution
 is unlikely to eliminate a need
 for $\Lambda $. 

\end{abstract}

\section{INTRODUCTION}
Two of the important developments in observational SN-research
in the last few years were to establish the long-suspected
correlation between the peak brightness of SNe~Ia and their
rate of decline by means of modern CCD photometry \cite{phillips87}, and the
exact distance calibrations provided by an 
 HST Key Project (e.g. \cite{saha97}). Both allowed a empirical determinations of $H_o$ with
unprecedented accuracy \cite{hamuy96}.
 Independent from these calibrations and empirical relations,
$H_o$ has been determined by   comparisons of detailed 
theoretical models for light curves and spectra with observations 
\cite{mh94,hk96,nugent96}.
 All determinations of the
Hubble constant are in good agreement with one another.
 Recently, the routine successful detection of SNe at 
large redshifts, z  \cite{perlmutter98,riess98},
has provided an exciting new tool to probe cosmology.  This work has 
provided results that are
consistent with a low matter density in the Universe and, most
intriguing of all, yielded hints for a positive cosmological constant.  
 The cosmological results rely on  empirical brightness-decline relations which are
calibrated locally. One of the main concerns are systematic, evolutionary effects in the
properties of SNe~Ia. Both from  theory and observations, there are some hints for the presents
of evolution. 

 Observational evidence include the finding of
 Branch et al. \cite{branch98} who have
shown that the mean peak brightness is dimmer in ellipticals 
than in spiral galaxies.  Wang, H\"oflich \& Wheeler \cite{wang97}  found that
the peak brightness in the outer region of spirals is 
similar to those found in ellipticals, but that in the central region both 
intrinsically brighter and dimmer SNe~Ia occur.
  This implies that the 
progenitor populations are more inhomogeneous in the inner parts
of spirals which contain both young and old progenitors.
 
From theory,  time evolution is expected to produce the following main effects: (a) a
lower   metallicity will decrease the time scale for  stellar evolution of individual stars
by about 20 \% from Pop I to Pop II stars \cite{schaller92} and,
consequently,
the  progenitor population which contributes to the SNe~Ia rate at any given
time.
 The stellar radius also shrinks. This will influence the statistics
of systems with mass overflow;
 (b) early on, we expect that systems with shorter life time will dominate the sample whereas,
today, old system are contributing which may have not occurred early on.
 In addition, some scenarios with a life time comparable to the age of the univers such as two
merging WDs may be absent a few Gyrs ago, but they may contribute today. 
 (c) The initial metallicity will determine the electron to nucleon fraction of
the outer layers  and hence affects the products of nuclear burning;
 (d) Systems with a shorter life time may dominate early on and, consequently,
the typical C/O ratio  of the central region of the WD will be reduced;
 (e) The properties of the interstellar medium may change;
 (f) The influence of Z on the structure of WDs may change, but this effect has remains very small (e.g. $2\% $
when comparing solar with 0.01 solar, \cite{hwt98})
(g) The distribution of C and O will depend on Z as it influences 
the `normal' stellar evolution and the properties of the C/O core \cite{schaller92}
(h) The metallicity will effect nuclear burning during the accretion phase of the progenitor 
\cite{nomoto2000}.
 
 Based on theoretical models described in \S III,
we want to study the possible effects of the change of the progenitor
and its metallicity on the light curves and spectra of SNe~Ia. Note that most of the results
have been obtained in serveral collaborations over the years (see acknowledgments).

\noindent
\section{Numerical Tools}
 
 Most of the results discussed in  the following sections are based on
our calculations which are consistent with
 the explosion, light curves and spectra. In some cases, the stellar evolution and the accretion on the WD
is treated in detail.
 
\noindent{\bf Stellar Evolution:}
The stellar models have been calculated using the evolutionary code
 Franec (e.g. \cite {scl97,cls98,dominguez2000}) or are provided by
Nomoto's  group \cite{umeda99}.
 Subsequently, the evolution of the C/O core is followed up by
accreting H/He rich material at  a given accretion rate on the core by solving the standard equations for stellar evolution
using a Henyey scheme.
 Nomoto's equation of state is used \cite{nomoto82}. Crystallization is neglected.
 For the energy transport, conduction \cite{itoh83},
convection  in the mixing length theory, and radiation is taken into account. Radiative opacities for free-free and
bound free transitions are approximated in Kramer's approximation  and by free electrons. A nuclear 
network of 35 species up to $^{24} Mg $ is used.

\noindent{\bf Hydrodynamics:}
The explosions are calculated using a one-dimensional radiation-hydro
code, including nuclear networks (\cite{hk96}            and
references therein).
 This code solves the hydrodynamical equations
explicitly by the piecewise parabolic method \cite{collela84}
and includes the solution of the frequency averaged radiation transport
implicitly via moment equations, expansion opacities (see below),  and a detailed
equation of state. Nuclear burning is taken into account using a network which has been tested in many 
explosive environments (see \cite{fkt96} and
references therein).

\noindent{\bf  Light Curves:}
Based on the explosion models, the subsequent expansion
 and  bolometric as well as monochromatic light
curves are calculated using a scheme recently developed, tested
and widely applied to  SN Ia (e.g. \cite{hkm93,hwt98}).
The code used in this phase is similar to that described above, but
 nuclear burning is neglected and
 $\gamma $ ray transport is included via a Monte Carlo scheme.
In order to allow for a more consistent treatment of scattering, we
solve both the (two lowest) time-dependent, frequency averaged radiation moment equations for the
radiation
energy and the radiation flux, and a total energy equation.
At each time step, we then use $T(r)$ to determine the
Eddington factors and mean opacities by solving the frequency-dependent
radiation transport equation in the comoving frame 
and integrate to obtain the frequency-averaged quantities.
About 1000 frequencies (in one 100 frequency groups) and
about 500 to 1000  depth points are used. 
 The averaged  opacities have
been calculated under the assumption
of local thermodynamic equilibrium. 
Both the monochromatic and mean opacities are calculated using the Sobolev
approximation.
The scattering, photon redistribution  and thermalization terms
used in the light curve opacity calculation are calibrated with NLTE
calculations using the formalism of the equivalent-two-level approach
\cite{h95}.

\noindent{\bf Spectral Calculations:}
Our  non-LTE code (e.g. \cite{h95,hwt98})
 solves the relativistic
radiation transport equations in comoving frame. 
The energetics of the SN
are calculated. The evolution of the spectrum is not subject to any
tuning or free parameters.
The  spectra are computed for various epochs using the chemical,
density and luminosity structure and $\gamma$-ray deposition resulting
from the light curve code 
providing a tight coupling between the explosion
model and the radiative transfer.  The effects of instantaneous energy
deposition by $\gamma$-rays, the stored energy (in the thermal bath and
in ionization) and the energy loss due to the adiabatic expansion are
taken into account.  The radiation transport equations are solved
consistently with the statistical equations and ionization due to
$\gamma $ radiation for the most important elements (C, O, Ne, Na, Mg,
Si, S, Ca, Fe, Co, Ni).  Besides 40,000 lines treated in full non-LTE, 
$\approx 10^{6}$ additional lines are included
assuming LTE-level populations and an equivalent-two-level approach for
the source functions.

\section{Models for Thermonuclear Supernovae }

\medskip\noindent
There is general agreement that SNIa result from the thermonuclear explosion of a
degenerate white dwarf (HF60). Within this general picture, three
classes of models have been considered: (1) An explosion of a CO-WD, with mass
close to the Chandrasekhar mass, which accretes mass through Roche-lobe overflow
from an evolved companion  star \cite{nomsug77}. The
explosion is then triggered by compressional heating near the WD center.  (2) An
explosion of a rotating configuration formed from the merging of  two low-mass WDs,
caused by the loss of angular momentum due to gravitational radiation
\cite{webbink84,ibentut84,paczy85}.  The explosion may also be
triggered near the center by the compression.  (3) Explosion of a low mass CO-WD
triggered by the detonation of a helium  layer \cite{nomoto80,ww86}.
Only the first two models appear to be viable.
The third, the sub-Chandrasechar WD model, has been ruled out on the basis of
predicted light curves and spectra \cite{h96b,nugent97}.

From the theoretical standpoint, one of the key questions is  how the flame propagates
through the WD. Several burning models have been proposed in the past,
including detonation \cite{arnett69,hw69},  deflagration 
\cite{ivanova74,nomoto76}, and the delayed
detonation model (DD) \cite{k91ab,ww94,yamaoka92}.
The DD-model assumes that burning starts as a subsonic
deflagration with a certain speed $S_{\rm def} < a_s$, where $a_s$ is the sound
speed,  and then undergoes a transition to a supersonic detonation.
  The detonation speed follows directly from the standard Hugoniot
relations. However,  due to the one-dimensional nature of current 
model, the speed of the subsonic deflagration and the moment of the transition to
a detonation are free parameters, or calibrated by 3-D calculations.
 The moment of deflagration-to-detonation
transition (DDT) is conveniently parameterized by introducing the transition
density, $\rho_{\rm tr}$, at which DDT happens. 
Recently,  significant progress has been made toward  a better 
understanding of the propagation of nuclear burning fronts during the deflagration phase.
The models provide good results during the phase when the Ryleigh-Taylor instabilites dominate but
the results depend sensitively on the assumed
sub-grid models at later phase of the flame propagation
 \cite{khokhlov95,khokhlov97,niemeyer95}, and the DDT-transition is not
understood quanitatively. Therefore, further constraints must come from the observations.

What we observe as a SN is not the explosion itself, but the light 
emitted from a rapidly expanding ejecta produced by the explosion. As  the
photosphere recedes, deeper layers of the ejecta become visible. A detailed 
analysis of multi-band light curves and spectra gives us the opportunity to
determine the density, velocity and composition structure of the ejecta, and to
constrain the physical model of the explosion. The major results of
one-dimensional modeling and comparison with observations are summarized
below.

\begin{figure}[h]
\psfig{figure=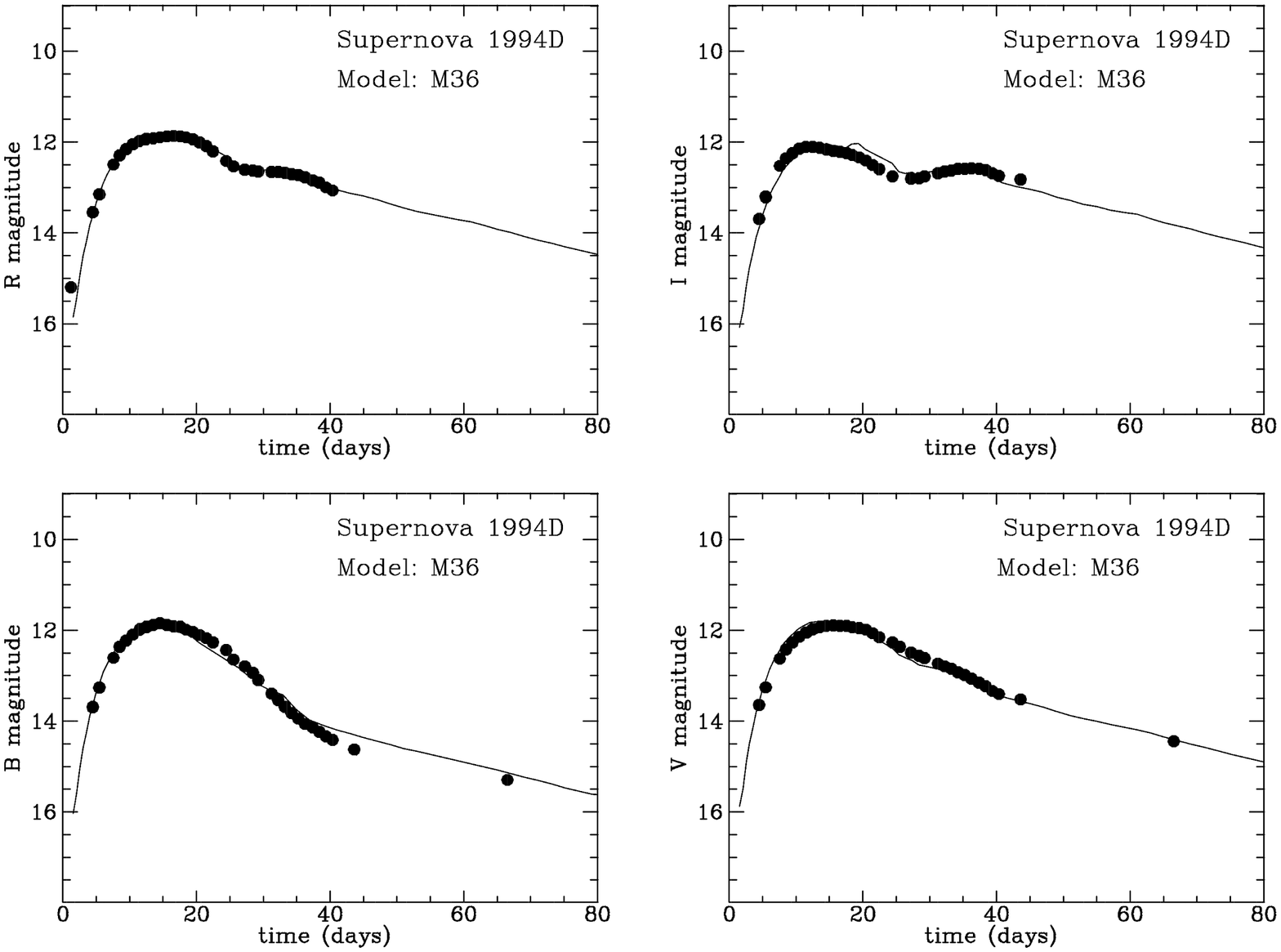,width=4.4cm,clip=,angle=270}
\psfig{figure=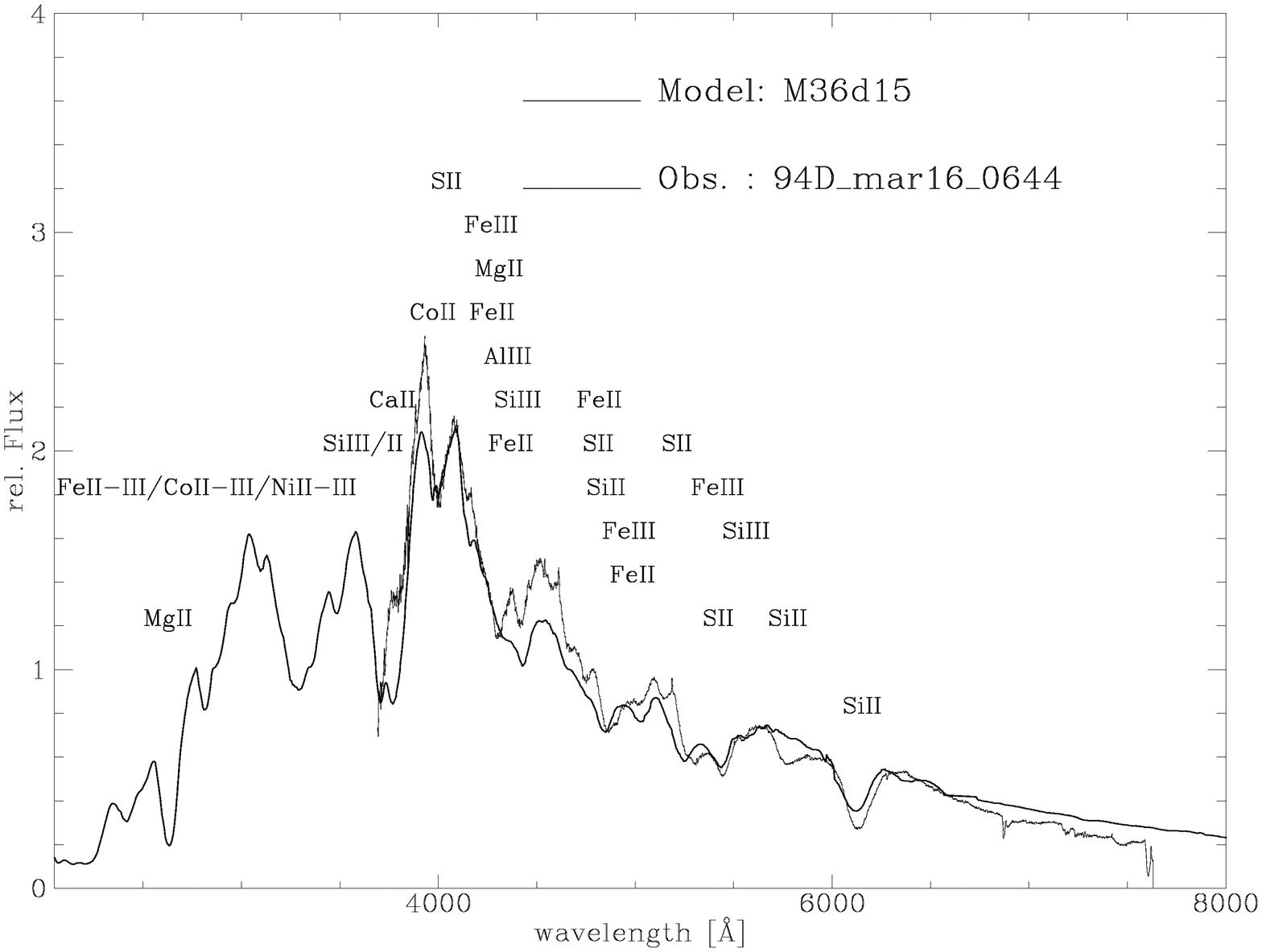,width=4.4cm,clip=,angle=270}
\caption {
 Observed LCs of SN 1994D in comparison with the theoretical LCs of a typical delayed
detonation model, and the corresponding synthetic spectrum around bolometric maximum in comparison 
to the observations of SN1994D at Mar. 16th (from H\"oflich 1995).} \label{SN94D}
\vskip -0.1cm
\end{figure}
\begin{figure}[h]
\vskip -0.4cm
\psfig{figure=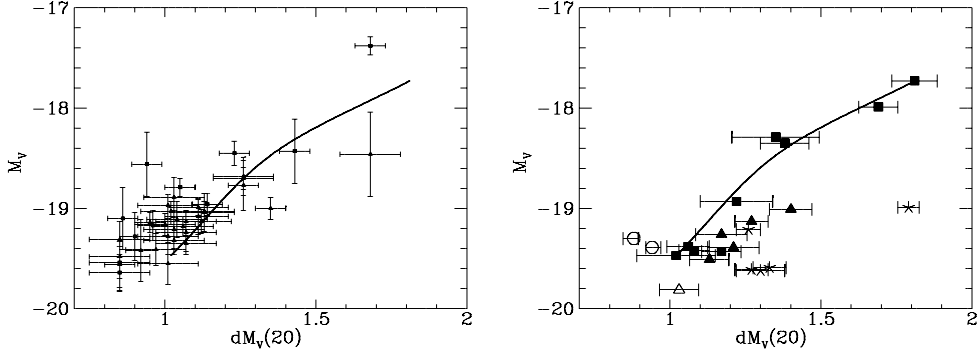,width=5.0cm,clip=,angle=270}
\vskip -0.0cm
\caption {(Left panel) Observed light curve maximum brightness -
decline rate relation.  M$_V$ is presented as a function of the decline from
maximum at 20 days. (Right panel)
 The predicted relation for an array of 
models of SNe~Ia representing delayed detonations (open triangles), 
pulsating delayed detonations (filled circles), 
merging models (open circles) and helium detonations (asterisks). For both the  
delayed detonation and merger scenarios  models are only  considered if they allow for a 
representation of some of the observed SNe~Ia (H\"oflich et al. 1996).} \label{dm}
\vskip -0.1cm
\end{figure}

 The best current explosion model, the delayed detonation
 and pulsating delayed detonation explosion models of
Chandrasekhar mass carbon-oxygen WDs  can account for the spectral and
light curve evolution  of both ``normally bright" and subluminous SNIa in the
optical and IR (see Fig. \ref{SN94D}, \cite{h95,hkw95,hk96,wheeler98}.
 Within this framework, we can understand
 for often used brightness decline relation.
 For a plausible range in  the transition density,
$\rho_{\rm tr}\simeq (1.5-2.5) \times 10^7$ g~cm$^{-3}$ \cite{hk96},
sufficient thermonuclear energy is generated  by burning nearly
the entire WD to provide the observed expansion velocities, a small spread in the explosion
energy \cite{h95}. The amount of $^{56}Ni$ varies 
between $\simeq 0.1 - 0.7$ M$_\odot$. The
 variation of $M_{\rm Ni}$ gives a range in maximum 
brightness that matches the observations (Fig. \ref{dm}).
 The models with less nickel are not
only dimmer, but are cooler  and have lower opacity, giving them redder, more
steeply declining  light curves, in agreement with the observations \cite{hkm93,hkwpsh96}.
The amount of $^{56}$Ni depends primarily on  $\rho_{tr}$, and
 to a much lesser extent on the assumed value of deflagration speed,
initial central density of the exploding star, and the initial chemical composition (ratio of
carbon to oxygen). This is the basis of why, to first approximation, SNIa appear to be a
one-parameter family. Nonetheless, variations of the other parameters also lead to some
variations of the predicted properties of SNIa, which indicate that the assumption of a
one-parameter family is not strictly valid.
We get a spread around the mean relation of $\approx 0.4^m$ which is consistent with the spread
based on the CTIO data  published by Hamuy et al. \cite{hamuy96} but larger than suggested by
recent observations
($\approx 0.12^m$ \cite{riess98}).
 This narrow a spread cannot be understood in the context of current models but
it may hint of an underlying coupling of the progenitor, the accretion rates and the
propagation of the burning front.

\begin{figure}[h]
\psfig{figure=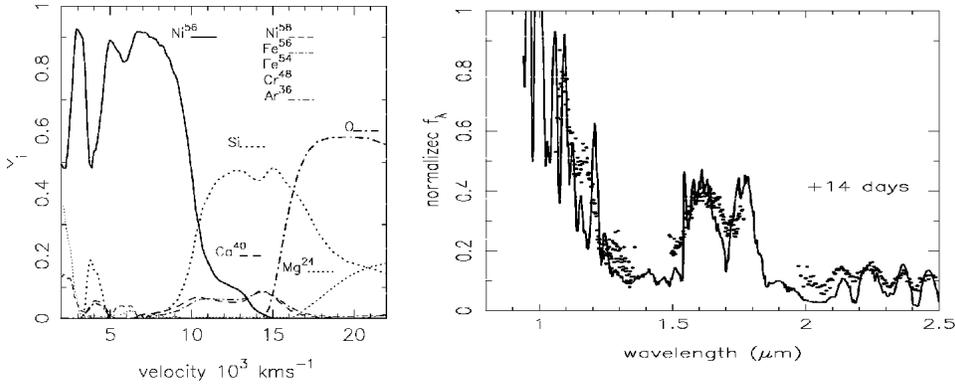,width=5.0cm,clip=,angle=270}
\caption {Final chemical structure of a  delayed detonation model as calculated in 1-D
in comparison and the theoretical IR-spectrum in the IR at about 2 weeks after
maximum light  in comparison with the observed spectrum of 1976g at about +10 days
by Bowers et al. (1997). The emission feature between 1.5 and 1.8 $\mu m$ is produced
by a large number of Fe, Co and Ni lines. The $^{56}Ni$ layers are optically thick
at these wavelengths. It appears as soon as the 'photosphere' at adjoining wavelengths
receds to velocities smaller than   the outer edge of the $^{56} Ni$ layers
(from  Wheeler, H\"oflich, Harkness and Spyromillo 1998). }\label{SN76g}
\end{figure}

 During the last 2 to 3 years, near IR spectra
 became available and provided a new insights.
Firstly, the comparison with observed IR spectra with model predictions was an
important test for the models  since there was no a priori
guarantee that a model that matched optical light curves would produce
satisfactory IR spectra (Fig. \ref{SN76g}, \cite{wheeler98} ). The broad feature between 1.5 and 1.8 $\mu m$
is produced by a large number of Fe, Co and Ni lines and indicates the size of the $^{56}Ni$ rich
regions.
 A very strong Mg II line at about 1.05 $\mu m$ at early times can be understood
as a natural consequence of burning in DD-models. Because Mg is produced
in the region of explosive carbon burning, the Doppler shift provides a
unique tool to determine the transition zone from explosive carbon to oxygen 
burning. Its  high, minimum velocity clearly demonstrates the need for a DDT transition
or a  deflagration speed well in excess of those of the deflagration model W7 \cite{nomoto84},
and it rules out strong mixing as a possible explanation for the presence of high velocity
Si frequently seen in 'normal' bright SNe~Ia \cite{bennetti89,fisher98}.

\begin{figure}[h]
\vskip -0.2cm
\psfig{figure=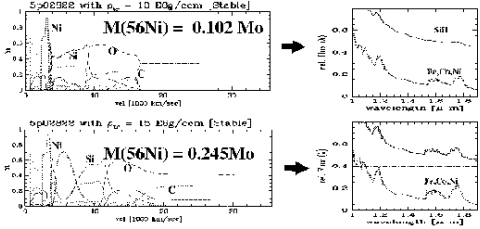,width=6.1cm,clip=,angle=270}
\caption {
Final chemical structure of two strongly subluminous  delayed detonation model and
 the corresponding, theoretical IR-spectrum   at  1 and 2 weeks after
maximum light. In the strongly subluminous SN1999bu, the 1.5 to 1.8 $\mu m$ feature was
not observed until about 2 weeks after maximum light. To be consistent with this observation,
 the abundance  $^{56}Ni$ must be less than $\approx 10 \%$ at layers with expansion velocities in
 excess of  $\approx  5000 km/sec$
Gerardy et al. (2000).}\label{subl}
\vskip -0.1cm
\end{figure}

Recently,  IR for the very subluminous SN1999bu by Gerardy et al. \cite{gerardy2000} shows
no Ni at layers with expansion velocities larger
than $\approx  5000 km/sec$.
Progress in  combustion
in SNeIa predicted (e.g. \cite{khokhlov95}) that, during the deflagration phase, some Ni bubbles may rise to
about half of the WD-radius corresponding to 0.4 $M_\odot$ which may be
inconsistent with results of the 1-D spectra (Fig.\ref{subl}) because, in subluminous SNe~Ia, most of the $^{56}Ni$
is produced during the deflagration phase. The solution to this
problem may  prove to be the Rosetta stone for  understanding  the influence of the preconditioning of the WD,
i.e. the progenitor  star, its rotation etc., and of  propagation of  burning fronts.
 Eventually, it may help to answer the question whether a SNIe~Ia is subluminous
because its special initial conditions influence the burning front, or whether
current models for the deflagration need to be modified.

\section{Evolutionary Effects}

  A change of the main sequence mass of the typical progenitor will change the size the region
of central He burning during the stellar evolution and, consequently, the size and C/O ratio
in the WD which, eventually, may explode as a SNe~Ia. The initial metallicity Z of the WD is
inherented from the molecular cloud from which the star has been formed.
 The influence of  Z  and C/O ratio on light curves and
spectra  has been studied for the example of a set of DD-models with  the basic properties
as follows:
central density of the WD, $\rho_c= 2.6 \times 10^9 g/cm^{-3}$,
$v_{burn}=\alpha *v_{sound}$ with $\alpha = 0.03$ during the deflagration phase and
a transition to detonation at $\rho_{tr}=  2.7 \times 10^7$.
The  quantities  of Table 1 
 in columns 2 to 5 and 6 to 9  are: 
  C/O ratio;     $R_Z$ the
Z relative to solar by mass; $E_{kin}$ kinetic energy (in
$10^{51}erg$);
$M_{Ni}$ mass of $^{56}Ni$ (in solar units).
The parameters  are close to those which
reproduce both the spectra and light curves reasonably well \cite{nomoto84,h95}.
\begin{table}
\begin{center}
\caption{ Basic parameters for 
the  delayed detonation models.
}
\vskip -0.1cm
\begin{tabular}{llllllllllll}
\+Model &  C/O & $R_Z$~ & $E_{kin}$~  & $M_{Ni}$~  && 
  Model &  C/O & $R_Z$~ & $E_{kin}$~  & $M_{Ni}$~  \\ 
\hline                     
\+DD21c   &     1/1 & 1/1  &  1.32  & 0.69  &&   DD25c   &     1/1 & 3/1  &  1.32  &  0.69  \\
\+DD23c   &     2/3 & 1/1 &  1.18  &   0.59  && DD26c   &   1/1 & 1/10  &  1.32  &   0.73  \\
\+DD24c   &     1/1 & 1/3  &  1.32  &  0.70  && DD27c   &  1/1 & 10/1  &  1.32  &   0.69  \\
\end{tabular}
\end{center}
\vskip -0.4cm
\end{table}
\begin{figure}[h]
\vskip -0.3cm
 \psfig{figure=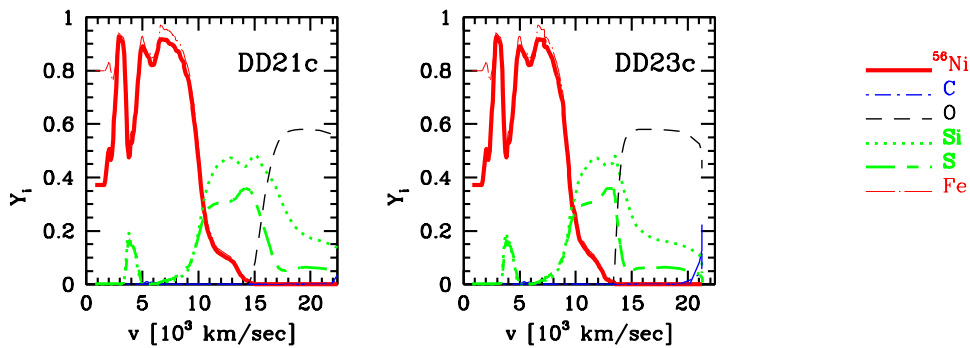,width=15.5cm,clip=,angle=360}
\caption{ Abundances as a function of the final expansion velocity for the
 delayed detonation models DD21c and DD23c.
Both the initial $^{56}Ni$ and the final Fe profiles are shown.}\label{dd21}
\vskip -0.1cm
\end{figure}
 
\noindent{\bf Direct influence of C/O:}
 As the C/O ratio of  the WD is decreased from 1 to 2/3, the explosion energy and
 the $^{56}Ni$ production are reduced and the Si-rich layers are more confined in
velocity space (Fig. \ref{dd21}).  A reduction of C/O by about 60 \% gives  slower
rise times by about 3 days and an increased luminosity at maximum light,
 a somewhat faster post-maximum decline and
a larger ratio between maximum light and the $^{56}Co$ tail (\cite{hwt98}, see also Fig.\ref{lc7}). The slight increase in luminosity
at maximum light is caused by the smaller expansion rate. Consequently, less
energy stored early on is wasted in expansion work, but contributes to the luminosity.
The smaller $^{56}Ni$ production causes the reduction of the luminosity later on.
 
 A reduction of the C/O ratio has a similar effect on the colors, light curve
shapes
and element distribution as a  reduction in the deflagration to detonation
transition
density but, for the same light curve shape, the absolute brightness
is larger for smaller C/O. Moreover, the kinetic energy is reduced by about 10 \% (Table 1) and,
consequently, the expansion velocity derived by the Doppler shift in the spectra becomes
smaller by about 5\% .
 An independent determination of  the initial C/O ratio and the transition
density
 is possible for local SN if detailed analyses of both the spectra and
light curves are performed simultaneously.

\begin{figure}[h]
\vskip -0.3cm
\psfig{figure=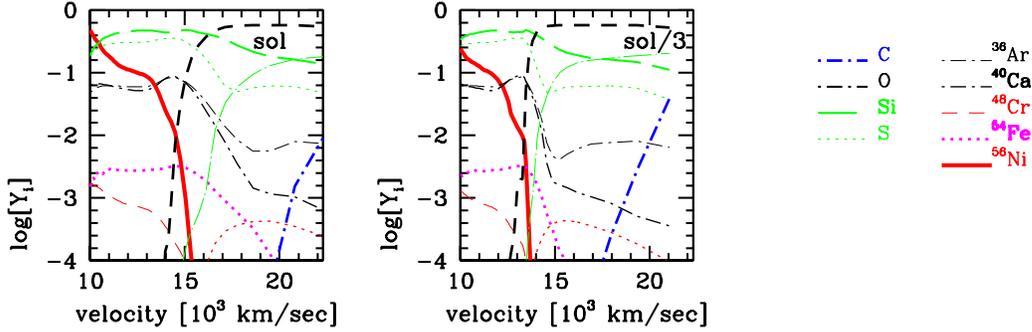,width=15.0cm,clip=,angle=360}
\caption{Abundances of different isotopes  as a function of the
expansion velocity for DD21c with initial abundances of solar and solar/3.}\label{iso}
\vskip -0.1cm
\end{figure}

\noindent{\bf Direct influence of the metallicity:}
 To test the influence of the metallicity for  nuclei beyond
Ca,  we
have constructed models
with parameters identical to DD21c but with initial metallicities
between 0.1 and 10 times solar (Table 1). The energy release,
the density and velocity  structure are virtually identical to that of DD21c and,
consequently, the bolometric (and also monochromatic) optical light curves are
rather insensitive.
The main influence of Z  is a slight increase of the $^{56}Ni$ mass with decreasing
Z due to a higher
$Y_e$.
 The reason is that Z mainly effects the initial CNO abundances
of a star.
 These are converted
during the pre-explosion stellar evolution to $^{14}$N in H-burning and via
$^{14}$N($\alpha,\gamma$)$^{18}$F($\beta ^+$)$^{18}$O($\alpha,\gamma$)$^{22}$Ne
to nuclei with
N=Z+2 in He-burning. The result is that increasing Z yields
a smaller proton to nucleon ratio $Y_e$
throughout the pre-explosive WD. Higher Z
and   smaller $Y_e$
lead to the production of more neutron-rich Fe group nuclei and less $^{56}Ni$ (Fig. \ref{iso}). 
 For lower Z and, thus,
 higher $Y_e$, some additional $^{56}Ni$
is produced  at the expense of $^{54}Fe$  and $^{58}Ni$
\cite{fkt1986}.                
 The temperature in the inner layers is sufficiently
high during the explosion that
 electron capture  determines $Y_e$. In those layers,
Z has no influence on the final burning product.
The main differences due to changes in Z
are in regions with expansion velocities in excess of $\approx $
12000 km/sec.
 Most remarkable is the change in the $^{54}$Fe production which is the
dominant contributor to the abundance  of iron group elements at these
velocities  since little cobalt has yet decayed near maximum light 
(Fig. \ref{iso}).
\begin{figure} [h]
\vskip  -.3cm
 \psfig{figure=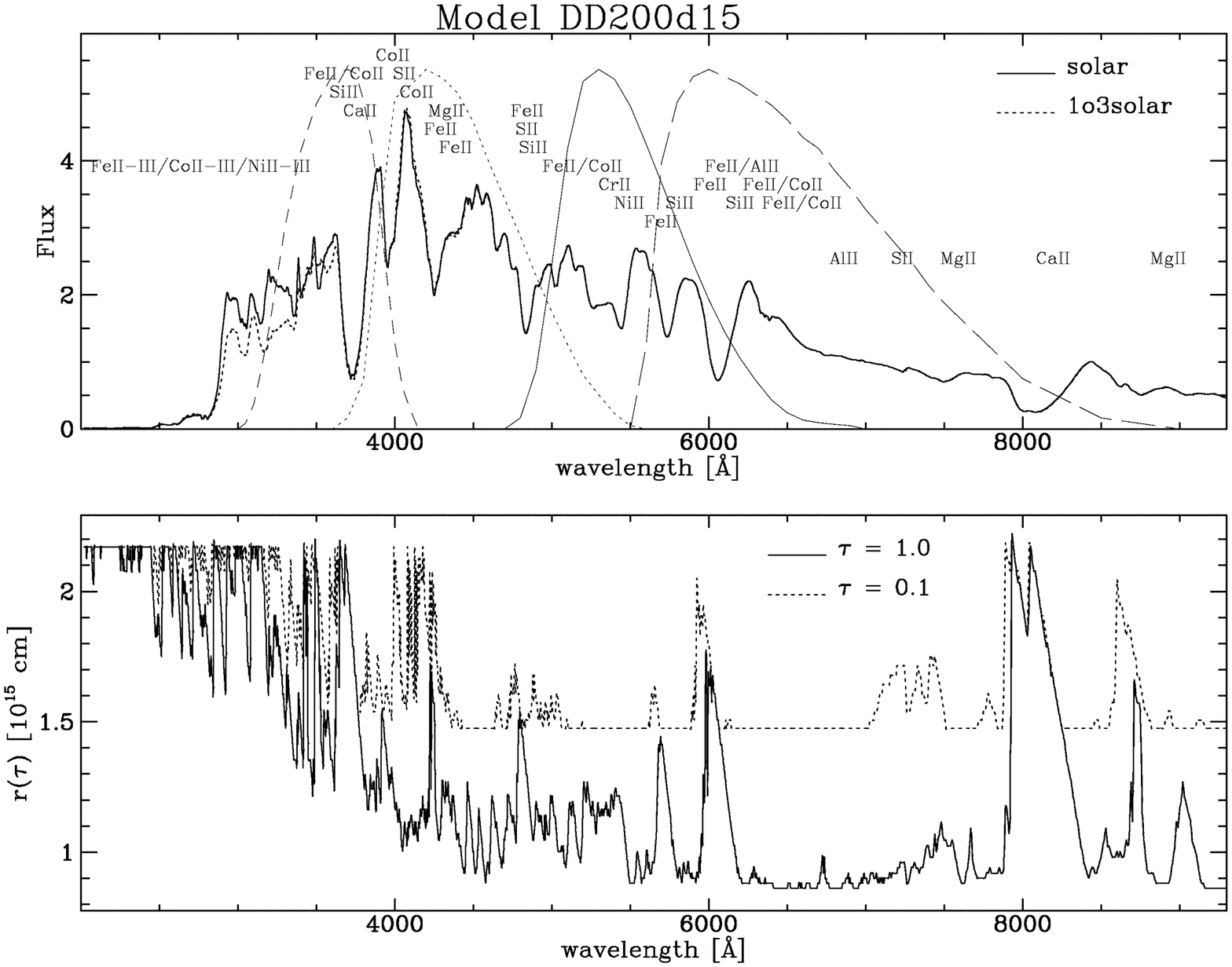,width=4.5cm,clip=,angle=270}
\caption{Comparison of synthetic NLTE spectra at maximum light for
 Z   solar (DD21c) and  1/3 solar (DD24c).
 The standard Johnson filter functions for UBV, and R are
also shown.}\label{specdd21}
\vskip  -.2cm
\end{figure}
 
 The initial WD composition has been found to have rather small effects on the overall LCs.
The $^{56}Ni$  production and hence the bolometric and monochromatic  optical and IR light curves 
differ only by a few
hundredths of a magnitude. This change is almost entirely due to the small change in the $^{56}Ni$
production and hardly  due to a change in the opacities because the diffusion time scales are governed
by the deeper layers where burning is complete.
However, the short wavelength part of the spectrum ($\lambda \leq 4000 \AA$) at maximum light is        
affected by a change in Z (Fig.\ref{specdd21}).
   This provides a direct test for Z of local SNe 
 and, thus, may give a powerful tool to unravel the nature (and lifetime) of  
SNe~Ia progenitors.

By 2 to 3 weeks after maximum, the spectra are completely insensitive
to the initial Z because the spectrum is formed in even deeper
layers where none of the important abundances are
effected by Z.
 Thus, for two similar bright SNe with similar
expansion velocities, a comparison between the spectral evolution can provide
a method to determine the difference  in Z, or it may be used to detect
evolutionary effects for distant SNe~Ia if high quality spectra are available.

\noindent{\bf Influence of the Stellar Evolution on the WD Structure:}
 Up to now, we have neglected the influence of the metallicity and the mass of the progenitor
on the structure of the initial WD for a given mass on the main sequence. Such dependencies may become of important if
SNe are observed at large distance.   On cosmological distance scales, Z
is expected to be correlated  with redshift.
 At the time of the explosion, the WD masses are close to the Chandrasekhar limit. The WD has
grown by accretion of H/He and subsequent burning from the mass of the central core of a star with
less than $\approx 7 M_\odot$ (Fig. 6). In the accreted layers, the C/O ratio is close to 1; however, the
initial mass of the C/O WD is determined by the stellar evolution.
The core mass depends on $M_{MS}$ of the progenitor and on Z.

 Here, we want to discuss the size of the metallicity effect using the example of a 7 $M_\odot $ model
with Z=0.02 and 0.004 (Fig.\ref{lc7}). Z mainly effects the convection during the stellar
Helium burning and, consequently, the size of the C/O core and the central C/O ratio.
 We note that the exact size of the effect and its sign depends  on  mass of the progenitor
at the zero age main sequence, and Z. Even for a given mass, the changes are 
not monotonic,  but may change sign from Pop I to Pop II to Pop III  \cite{dominguez2000,umeda99}.
 In addition, the tendency depends sensitively on the assumed physics such as the
$^{12}C(\alpha, \gamma )^{16}O$-rate (e.g. \cite{scl97}).
Therefore, our example can serve as a guide to estimate the size of this effect.
\begin{figure}  [h]
 \psfig{figure=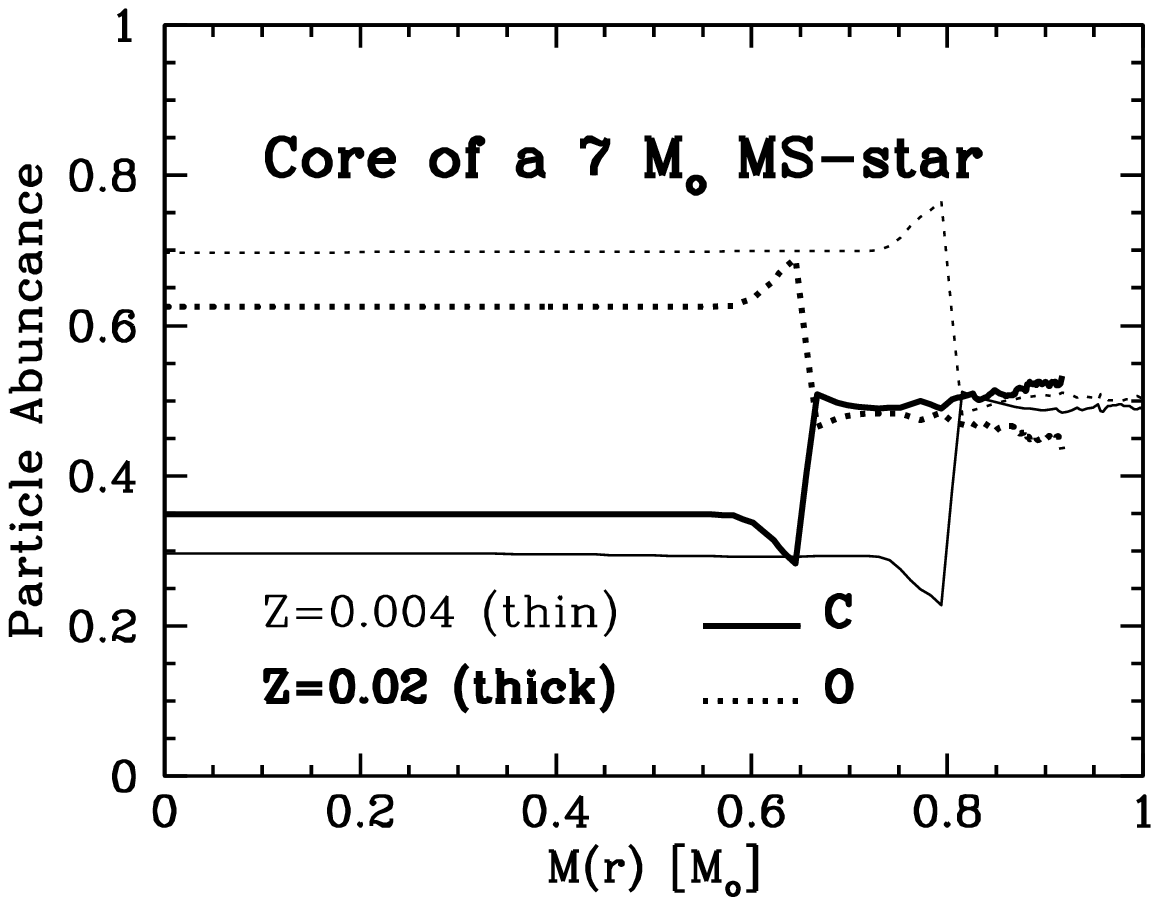,width=5.0cm,clip=,angle=270}
\vskip -5.cm
 \psfig{figure=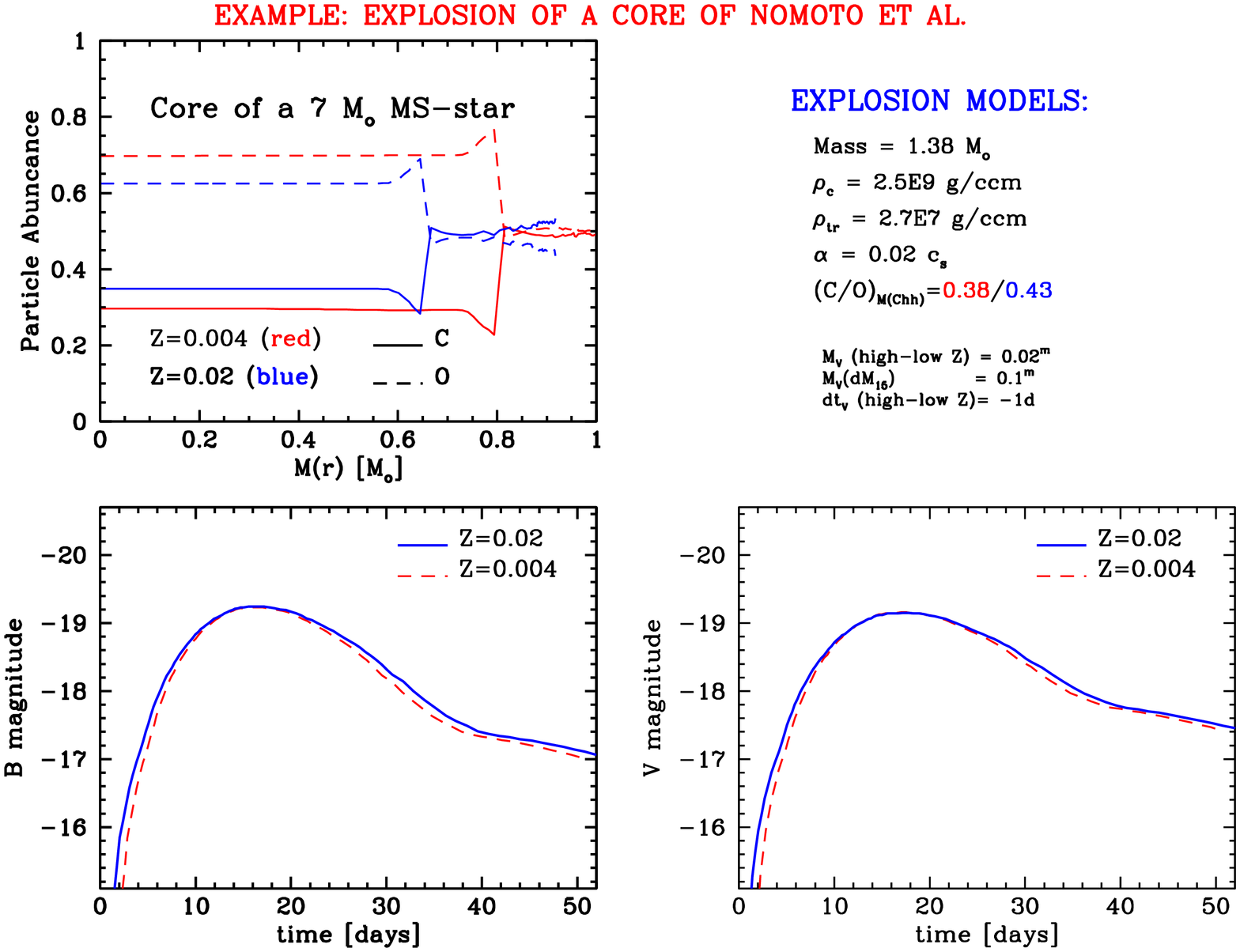,width=5.0cm,clip=,angle=270}
\caption{Comparison of the final chemical profile of the C/O core of a star with
 $M_{MS}=7 M_\odot$ (right) with Z=0.02 and 0.004, and the
 corresponding V light curves (left) of  delayed detonation
models with identical central density and descriptions for the burning front (see text).}\label{lc7}
\vskip -0.3cm
\end{figure}

 In agreement with above, the total C/O mass determines the explosion energy and, consequently,
has the main effect on the light curves.
After accretion on the initial core $M_C$, the total mass fraction of $M_C/M_{Ch}$ is 0.75 and 0.61 for the 
models with Z=0.04 and 0.02 respectively. 
 At the time of the explosion, $\rho_c=2.4 \times 10^9 g/cm^{-3}$. For the burning, 
 $\alpha =0.02$ and  $\rho_{tr}=2.4~10^7 g~cm^{-3}$).

Monochromatic LCs  are shown in Fig. \ref{lc7}. As expected from the last section,
 the main effect on the light curves is caused by the different expansion ratio determined by
the integrated C/O ratio. The change in the maximum brightness remains 
small ($M_V (Z=0.02) - M_V(Z=0.04)= -0.02^m$) and the rise times are different by about 1 day
($t_V((Z=0.02)=17.4d$ vs. $t_V(Z=0.04)=16.6 d$). The most significant effect is a 
steeper decline ratio and a reduced $^{56}Ni$ production for the model with solar Z.
This is mainly due to the slower expansion ratio. This translates into a systematic offset of $\approx 0.1^m$ in 
the maximum brightness decline ratio \cite{hamuy96}. Using either the streching method 
or the LCS-method gives similar offsets. 
 
Worth noting  is the following trend:
For realistic cores, the mean $M_C/M_{O}$ tends to 
be smaller than the canonical value of 1
used in all calculations prior to 1998 (e.g. \cite{nomoto84,ww94,hk96}).                     
 Consequently, as a general trend, the rise times are about 
1-3 days slower in all models based on the detailed WD structure  compared to the
 models published prior to 1996 (\cite{hk96}, and references therein).

\noindent{\sl Acknowledgments:} Most of the results reviewed here have been obtained in collaborations
with collegues  too numerous to be listed all as coauthors. Therefore,
 I would like to thank here, in particular, A. Chieffi, I. Dominguez, R. Fesen, C. Gerardy,
A. Khokhlov, M. Limongi, E. M\"uller, K. Nomoto, J. Spyromillo, Y.Stein,
 O. Straniero, F.K. Thielemann, H.Umeda, L. Wang \& J.C. Wheeler.
 This work is supported in part by NASA Grant                  
LSTA-98-022.

\end{document}